\documentclass[epj-spec]{svjour}
\usepackage{graphics,amssymb}
\usepackage[utf8]{inputenc}

\begin{document}
\title{Heat conduction in low-dimensional quantum magnets}
%\subtitle{Do you have a subtitle?\\ If so, write it here}
\author{Christian Hess\inst{1}\fnmsep\thanks{\email{c.hess@ifw-dresden.de}}}
\institute{Leibniz-Institute for Solid State and Materials Research, IFW-Dresden,
01171 Dresden, Germany}
\abstract{Transport properties provide important information about the mobility, elastic and inelastic of scattering of excitations in solids. Heat transport is well understood for phonons and electrons, but little is known about heat transport by magnetic excitations. Very recently, large and unusual magnetic heat conductivities were discovered in low-dimensional quantum magnets. This article summarizes experimental results for the magnetic thermal conductivity $\kappa_\mathrm{mag}$ of several compounds which are good representations of different low-dimensional quantum spin models, i.e. arrangements of S=1/2 spins in the form of two-dimensional (2D) square lattices and one-dimensional (1D) structures such as chains and two-leg ladders. Remarkable properties of $\kappa_\mathrm{mag}$ have been discovered: It often dwarfs the usual phonon thermal conductivity and allows the identification and analysis of different scattering mechanisms of the relevant magnetic excitations.
} %end of abstract
\maketitle
\section{Introduction}
\label{intro}

Heat transport by magnetic excitations was originally predicted in 1936 \cite{Froehlich36}. However, it took almost 30 years until the first convincing experimental evidence for magnetic heat transport by classical spin waves was found in ferrimagnetic yttrium-iron-garnet (YIG) \cite{Luethi62,Douglass63,Rives69,Walton73}. In principle, the analysis of this magnon heat conductivity should yield valuable information about the excitation and scattering of magnons (e.g. off defects, phonons, and electrons) as is the case for the well-understood phononic and electronic thermal conduction \cite{Berman}. However, most of the early experiments on YIG and following experiments on other materials \cite{Gorter69,Lang77,Coenen77} were restricted to magnetically ordered phases at very low temperature ($T<10$~K). The first signature of magnetic heat transport at higher temperatures ($T>50$~K) was observed for the one-dimensional quantum antiferromagnet $\rm KCuF_3$ \cite{Hirakawa75}. However, only the recent theoretical prediction of dissipationless heat conduction in one-dimensional antiferromagnetic Heisenberg chains \cite{Zotos97,Zotos99} and the discovery of huge magnetic contributions in the quantum spin ladder material $\rm Sr_{14}Cu_{24}O_{41}$ \cite{Kudo99,Sologubenko00,Hess01} triggered intense research on the heat transport of low-dimensional quantum spin systems \cite{Zotos97,Zotos99,Kudo99,Sologubenko00,Hess01,Hess02,Hess03,Hess04a,Hess04,Hess05,Ribeiro05,Hess06,Hess07,Hess07a,Sologubenko00a,Sologubenko01,Kudo01,Kudo03,Ando98,Vasilev98,Hofmann02,Hofmann03,Sologubenko03,Sologubenko03a,Heidrich02,Heidrich03,Orignac03,Alvarez02,Heidrich04,Gros04} \cite{Saito03,Saito03a,Shimshoni03,Klumper02,Sakai03,Zotos04,Karadamoglou04,Prelovsek04,Louis06,Li02,Rozhkov05,Chernyshev05,Jung06,Kordonis06,Boulat06}.
Over the course of this work, more and more cases for low dimensional magnetic heat conduction were observed in various materials.
Today, the clearest experimental examples of low dimensional magnetic heat conduction are found in copper oxide (cuprate) systems. The overview of the experimental research status on low dimensional magnetic heat conduction provided in this article therefore focuses on these compounds.

\begin{figure}
\resizebox{\columnwidth}{!}{
\includegraphics{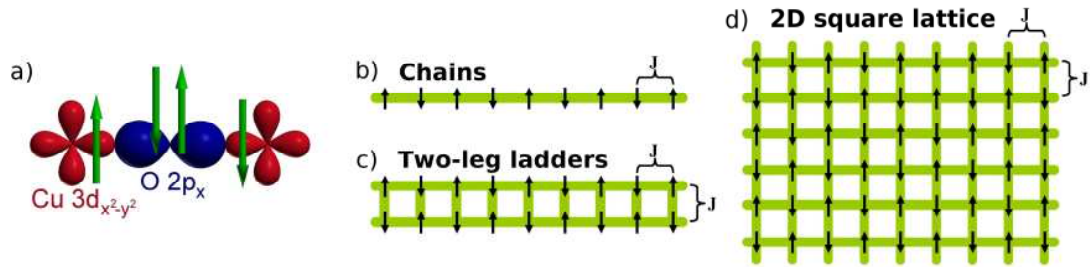} }
\caption{Illustration of low-dimensional spin structures: (a) a spin chain, (b) a two-leg spin ladder, and (c) a two-dimensional square lattice. Arrows represent localized $S=1/2$ spin and shaded bars symbolize strong antiferromagnetic exchange between them. (d) Schematic illustration of the underlying chemical building block giving rise to the spins and their interaction. Only the relevant Cu $3d_{x^2-y^2}$ and O $2p_x$ orbitals are indicated. Arrows represent the spins of the electrons involved.}
%\caption{Please write your figure caption here.}
\label{fig:1}       % Give a unique label
\end{figure}

Particular examples from the plethora of possible spin structures in cuprate systems are spin arrangements in the geometrical form of chains, so-called two-leg ladders, and square lattices with a strong antiferromagnetic Heisenberg exchange ($J\approx 1500-2000$~K) between nearest neighbor spins. Sketches of such spin arrangements are shown in Fig.~\ref{fig:1}a-c.
These low-dimensional spin structures usually arise from similar low dimensional {\rm Cu-O} structures, in which the antiferromagnetic exchange
originates from straight {\rm Cu-O-Cu} bonds (bonding angle: $180^\circ$) as depicted in Fig.\ref{fig:1}d. All these systems are
based on $\rm Cu^{2+}$-ions and therefore represent $S=1/2$ systems with a strong quantum nature.

Good examples for materials containing $S=1/2$ Heisenberg chains as depicted in Fig.~\ref{fig:1}a are given by the compounds $\rm CaCu_2O_3$, $\rm SrCuO_2$ and $\rm Sr_2CuO_3$, where straight Cu-O-Cu bonds and hence a strong antiferromagnetic exchange only exist along one particular crystallographic direction; the magnetic exchange perpendicular to this direction is much weaker \cite{Kiryukhin01,Motoyama96}. In the $\rm (Sr,Ca,La)_{14}Cu_{24}O_{41}$ family of compounds, parallel pairs of such chains are coupled to each other via bridging O-ions, producing in straight Cu-O-Cu bonds perpendicular to the chain direction. The resulting magnetic
interchain coupling perpendicular to the chain direction $J_\perp$ is of a similar magnitude to the intrachain coupling, i.e. $J_\perp\approx J$ \cite{Dagotto99}. This spin ensemble is a so-called two-leg spin ladder, the ladder legs being formed by the two chains and the ladder rungs arising from the Cu-O-Cu bonds which connect the chains (cf. Fig.~\ref{fig:1}b). Following this concept, ladder structures with more legs can in principle be created by coupling more chains to the structure; eventually this would lead to a two dimensional Heisenberg antiferromagnet on a square lattice (2D-HAF) in the infinite limit. A good realization of a 2D-HAF with $S=1/2$ is given by $\rm La_2CuO_4$ and other related antiferromagnetic parent compounds of high-temperature superconductors.

The corresponding low-dimensional quantum spin models are characterised by very peculiar ground state properties and elementary excitations, which vary strongly from system to system. Spin-spin correlations of homogeneous spin chains, for example, are quasi-long range in the ground state and decay algebraically with distance between the spins \cite{Kluemper93}. The elementary excitations, so-called spinons, are gapless and carry a spin $S=1/2$ \cite{Faddeev81}. In contrast to this, the ground state of a two-leg ladder is a
spin liquid, i.e. the spin-spin correlations are short range and decay exponentially as a function
of distance. The elementary excitations are $S=1$ particles (usually called magnons or triplons) and possess a spin gap $\Delta$ ($\Delta/k_B\approx400$~K in the case of the systems discussed here) \cite{Dagotto99}. Finally, the ground state of the 2D-HAF is a N\'{e}el state with strongly reduced sublattice magnetization, which only exists at temperature $T = 0$ \cite{Manousakis91}. In this case the elementary excitations are well described using a spin wave framework. However, it should be noted
that alternative descriptions have been discussed \cite{Coldea01,Sandvik01,Ho01}.

In the case of hole doping, all these model systems yield interesting and exotic properties. A Luttinger
liquid forms in hole-doped spin chains, i.e. electronic excitations decay into collective
excitations of holes (holons) and spins (spinons). This phenomenon is usually called spin-charge
separation. Radically different properties have been predicted for two-leg spin ladders:
superconductivity competing with a charge ordered ground state is expected in this case
\cite{Dagotto92,Dagotto96}. Finally, hole doping has great importance in the case of the 2D-HAF
as is evident from the high temperature superconductivity which is observed in such systems. Note that controlled hole doping of $S=1/2$ chains and the observation of spin-charge separation signatures has not yet been achieved experimentally, whereas charge ordering and superconductivity are prominent experimental features of hole-doped spin ladder and 2D-HAF materials.

Concerning heat transport, little is known for all these model systems. Often the attention in theoretical works is focussed on the possibility of ballistic magnetic heat transport in 1D-systems: in integrable models like the $XXZ$ Heisenberg spin chain the Hamiltonian and the thermal current operator commute, i.e. once a thermal current is established in such a system it will never decay \cite{Zotos97}. In other words, the thermal resistance vanishes and the magnetic thermal conductivity $\kappa_\mathrm{mag}$ diverges. While such surprising properties are well established for integrable spin models \cite{Zotos97}, ballistic heat transport in non-integrable quasi 1D-systems (e.g. two-leg spin ladders) is currently a subject of intense discussion \cite{Alvarez02,Zotos04,Heidrich04,Boulat06}. However, in real materials scattering processes involving defects and other quasiparticles such as phonons and charge carriers must play an important role and render $\kappa_\mathrm{mag}$ finite in all cases \cite{Shimshoni03}. The analysis of $\kappa_\mathrm{mag}$ should hence provide further insight into the nature of these scattering processes and the dissipation of magnetic heat currents.

\section{Experimental signatures of magnetic heat conduction}
\begin{figure}
\resizebox{\columnwidth}{!}{
\includegraphics{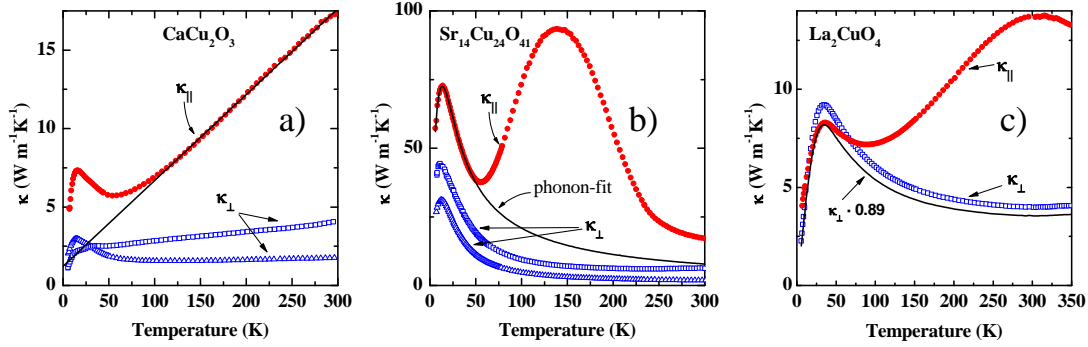} }
\caption{\label{kappadat}Anisotropic thermal conductivity of various low-dimensional spin materials as a function of temperature: (a) the spin chain compound $\rm CaCu_2O_3$, (b) the two-leg spin ladder material $\rm Sr_{14}Cu_{24}O_{41}$, and (c) the 2D-HAF as realized in $\rm La_2CuO_4$. Filled and open symbols represent $\kappa_\Vert$ and $\kappa_\bot$ of the materials. The solid line in (a) represents a linear fit to the data in the range $T\gtrsim 100$~K. The axis intercept of its extrapolation towards $T=0$ is an approximation of $\kappa_\mathrm{ph}$ in the fit range. Solid lines in (b) and (c) represent estimations for the phonon background. From Refs.~\cite{Hess07,Hess01,Hess03}.}
\label{kappa_raw}
\end{figure}
Fig.~\ref{kappa_raw} shows experimental results \cite{Sologubenko00,Hess01,Hess03,Hess07} for the thermal conductivity $\kappa$ of $\rm CaCu_2O_3$, $\rm Sr_{14}Cu_{24}O_{41}$ and $\rm La_2CuO_4$ which are good experimental representations of $S=1/2$ isotropic antiferromagnetic Heisenberg spin chains, two-leg spin ladders and the 2D-HAF, respectively. The experimental thermal transport properties of all these electrically insulating\footnote{Despite the spin ladder compound $\rm Sr_{14-x}Ca_xCu_{24}O_{41}$ being intrinsically hole doped, electronic contributions to $\kappa$ are negligible.} are  materials are intriguing: when $\kappa$ is measured perpendicular to the low dimensional structure ($\kappa_\bot$), i.e. a direction along which the magnetic coupling is negligible, the $T$-dependence of ordinary phonon thermal conductivity \cite{Berman} $\kappa_\mathrm{ph}$ is found for all three materials: $\kappa_\bot$ exhibits a peak at low temperature $T\approx20$~K, which is followed by a continuous decrease as $T$ rises further. Note the exception in $\rm CaCu_2O_3$, where one component of $\kappa_\bot$ increases monotonically with rising $T$. Here, a strong suppression of $\kappa_\mathrm{ph}$ due to disorder and possible contributions from optical phonons could give rise to the observed $T$-dependence \cite{Hess07}. The situation is completely different for $\kappa$ parallel to the low-dimensional structures, i.e. along the directions with large $J$ ($\kappa_\Vert$). Again, a phononic low-T peak is observed. However, $\kappa_\Vert$ evolves very differently at higher $T$. At $T\gtrsim 75$~K$, \kappa_\Vert$ strongly increases upon heating and exhibits a peak for $\rm Sr_{14}Cu_{24}O_{41}$ and $\rm La_2CuO_4$ at 140~K and 310~K, respectively, while for $\rm CaCu_2O_3$ the increase continues up to the highest temperature measured. In these three cases, the remarkable anisotropy of $\kappa$ is the qualitative evidence for large magnetic contributions to $\kappa_\Vert$, i.e. for magnetic heat conductivity $\kappa_\mathrm{mag}$. A similar anisotropy is also present for $\kappa_\Vert$ of the spin chain materials $\rm (Sr,Ca)CuO_2$ and $\rm Sr_2CuO_3$ which led to the conclusion that spinon heat transport is also present in these materials \cite{Sologubenko00a,Sologubenko01,Ribeiro05}. However, as will be discussed in Section~\ref{other}, a high-$T$ peak is absent in those cases which renders a quantitative analysis of phononic and magnetic contributions to $\kappa_\Vert$ more difficult. However, a clear separation of $\kappa_\Vert$ into magnetic and phononic parts ($\kappa_\mathrm{mag}$ and $\kappa_\mathrm{ph}$) is possible for the three cases shown in Fig.~\ref{kappa_raw} where the strong features of $\kappa_\mathrm{mag}$ appear at a much higher $T$-scale than the low-$T$ phonon peak. In the following sections we will therefore focus on these cases and examine what can be learned.

\subsection{Extraction of magnetic contributions}
\begin{figure}
\resizebox{\columnwidth}{!}{\includegraphics{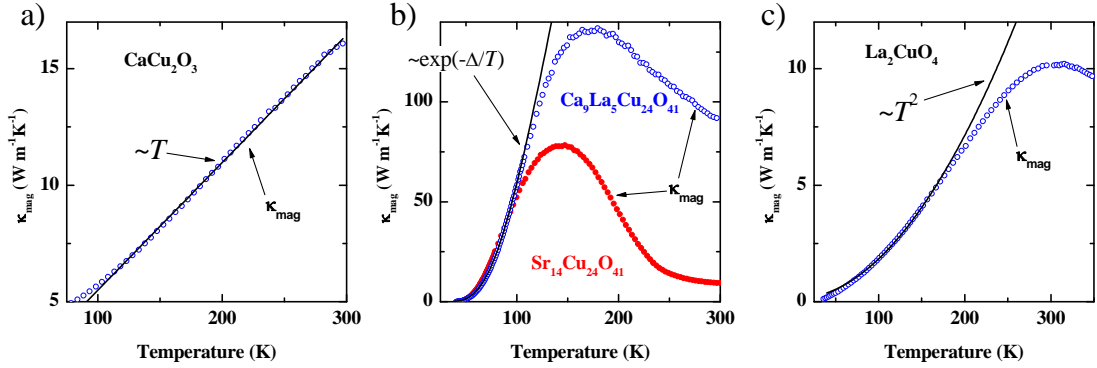}}
\caption{Estimated magnetic thermal conductivity as a function of temperature of (a) the spin chain compound $\rm CaCu_2O_3$, (b) the two-leg spin ladder compounds $\rm Sr_{14}Cu_{24}O_{41}$ ($\bullet$) and $\rm Ca_9La_5Cu_{24}O_{41}$ ($\circ$), and (c) the 2D-HAF  $\rm La_2CuO_4$. The solid lines represent fits according to the respective expressions for $\kappa_\mathrm{mag}$ in selected $T$-ranges (see text) with the approximate $T$-dependence indicated. From Refs.~\cite{Hess07,Hess01,Hess03}.}
\label{low_t}
\end{figure}

In order to obtain the magnitude and $T$-dependence of $\kappa_\mathrm{mag}$, it is essential to accurately estimate $\kappa_\mathrm{ph}$ and subtract it from the total $\kappa_\Vert$. In the case of $\rm La_2CuO_4$ and $\rm Sr_{14}Cu_{24}O_{41}$ this can be performed in a convenient way since the magnetic contributions are expected to be negligibly small in comparison to $\kappa_\mathrm{ph}$ for $T\lesssim40$~K, i.e. in the vicinity of the phononic peak. This is due to the expected $T$-dependence of $\kappa_\mathrm{mag}$ being approximately $\propto T^2$ and $\propto\exp{-\Delta/T}$ for the 2D-HAF and the spin ladders, respectively (cf. also the analysis further below). We may hence estimate $\kappa_\Vert\approx\kappa_\mathrm{ph}$ at $T\lesssim40$~K and to extrapolate the low temperature $\kappa_\mathrm{ph}$ to higher $T$. The thus estimated $\kappa_\mathrm{ph}$ is indicated in Fig.~\ref{kappa_raw}b and \ref{kappa_raw}c as solid lines. For details of the procedure the reader is referred to Refs.~\cite{Sologubenko00,Hess01,Hess03}. In the case of $\rm CaCu_2O_3$ this procedure is not applicable since $\kappa_\mathrm{mag}\propto T$ is expected for a $S=1/2$ Heisenberg chain hence providing significant contributions to $\kappa_\Vert$ even at low $T$. However, in the present case of $\rm CaCu_2O_3$ the situation is quite fortunate, since $\kappa_\Vert\gg\kappa_\bot\approx\kappa_\mathrm{ph}\approx \mathrm{const}$ at $T\gtrsim100$~K which allows the extraction of $\kappa_\mathrm{mag}$ from the total $\kappa_\Vert$ at temperatures higher than 100~K simply by subtracting a constant value \cite{Hess07}. For all three cases the conjectured $\kappa_\mathrm{mag}$ are shown in Fig.~\ref{low_t}. The figure also shows $\kappa_\mathrm{mag}$ of $\rm La_5Ca_9Cu_{24}O_{41}$, which belongs to the same family of two-leg spin ladder compounds as $\rm Sr_{14}Cu_{24}O_{41}$ but differs by a lower content of charge carriers in the ladders (cf. discussion in Section~\ref{high_T}).

\section{Analysis of magnetic heat conductivity}
We start our analysis by considering the qualitative $T$-dependence of $\kappa_\mathrm{mag}$, which comprises a simple peak structure ($\rm La_2CuO_4$ and $\rm Sr_{14}Cu_{24}O_{41}$) and a monotonic increase ($\rm CaCu_2O_3$) in the studied range $T=100$-350~K, where the latter may be regarded as the low temperature edge of a peak. A peak structure is very common for the thermal conductivity $\kappa$ of any kind of heat carrying particle, such as phonons or electrons \cite{Berman}. In the following we will investigate whether the underlying physics can be applied to magnetic excitations as well. The basic physics which determines the $T$-dependence of $\kappa$ can be inferred from the kinetic estimate
\begin{equation}
 \kappa=\frac{1}{d}\frac{1}{(2\pi)^d}\int c_{\bf k}v_{\bf k}l_{\bf k}d{\bf k}, \label{kinetic}
\end{equation}
with $d$ the dimensionality of the considered system, $c_{\bf k}=\frac{d}{dT}\epsilon_{\bf k}n_{\bf k}$ the specific heat ($\epsilon_{\bf k}$ and $n_{\bf k}$ are the energy and the statistical occupation function of the mode $\bf k$), $v_{\bf k}$ the velocity and $l_{\bf k}$ the mean free path of a particle with wave vector ${\bf k}$. At low $T$ only a few particles are excited and contribute to the heat transport. Often scattering processes are rare and $l_{\bf k}$ is a slowly varying function of momentum in the relevant energy range. In this situation, the low-$T$ increase of $\kappa$ is (a) characteristic of the excitation of the heat carrying particle (reflecting the $T$-dependence of the specific heat if $v_{\bf k}$ is momentum independent) and (b) proportional to the mean free path $l\approxeq l_{\bf k}$.
At higher $T$ the momentum-dependent scattering becomes more important and leads to a decrease of the mean free path and hence to a decrease of $\kappa$. Normally this decrease is characteristic of the relevant scattering mechanisms and allows an advanced analysis.   

The application of Eq.~\ref{kinetic} for the case of 1D and 2D magnetic systems leads to the general result 
\begin{equation}
 \kappa_\mathrm{mag}(T)\propto l_\mathrm{mag} f(T)~,\label{kmag_simple} 
\end{equation}
where $l_\mathrm{mag}(T)$ is a general magnetic mean free path based on the approximation $l_\mathrm{mag}\equiv l_{\bf k}$. For $k_BT\ll J$ this assumption is justified because the heat carrying excitations exist in significant numbers only in the vicinity of the band minima, i.e. a very small fraction of the Brillouin zone. This requirement is always fulfilled for all systems discussed in this article since $J\approx 1500$-2000~K and the experimental data only extend over temperatures $T < 350$~K.
The function $f(T)$ depends on temperature in a manner which is characteristic of the considered spin system.\footnote{For all three different types of systems the reader is referred to the original literature \cite{Hess01,Hess03,Sologubenko01,Hess07} for the derivation of the respective  $f(T)$. Within this article, we specify only the final results for $\kappa_\mathrm{mag}$.} 
In particular, for a gapless $S=1/2$ Heisenberg chain Eq.~\ref{kinetic} yields \cite{Sologubenko01}
\begin{equation}
 \kappa_\mathrm{mag}=\frac{2n_s{k_B}^2}{\pi\hbar}l_\mathrm{mag}T\int_0^\frac{J\pi}{2k_BT}x^2\frac{\exp(x)}{(\exp(x)+1)^2}dx~,
\end{equation}
where $n_s$ is a geometrical factor that counts the number of chains per unit area. At low temperatures $k_BT\ll J$ the integral is only weakly temperature dependent and approaches the constant value $\pi^2/6$ for $T\rightarrow0$. Note that the condition $k_BT\ll J$  holds even at room temperature (up to $T\lesssim 0.15 J/k_B\approx300$~K), i.e. the experimental data always represent the low temperature behaviour of $\kappa_\mathrm{mag}$ \cite{Sologubenko01,Hess07} and the upper boundary of the integral may be set to infinity.
The corresponding expression for a gapped two-leg ladder is \cite{Hess01}
\begin{equation}\label{kappalad}
\kappa_\mathrm{mag}=\frac{3 n_s {k_B}^2 }{\pi\hbar}l_\mathrm{mag}T \int_\frac{\Delta}{k_BT}^\infty
x^2 \frac{\exp(x)}{(\exp(x)+3)^2}dx~,
\end{equation}
where $n_s$ is now the number of ladders per unit area. 
At low temperatures $k_BT\ll \Delta<J$ one might approximate $f(T)\propto \exp(-\Delta/(k_BT))$.
Finally, for a 2D-HAF one finds, accounting for both magnon branches \cite{Hess03},
\begin{equation}\label{fit2d}
\kappa_\mathrm{mag}=\sum_{i=1,2}\frac{n_p{k_B}^3}{4\pi\hbar^2v_0} \,l_\mathrm{mag} \, T^2 
\int_\frac{\Delta_i}{k_BT}^\infty x^2\sqrt{x^2-x_{\mathrm{0,i}}^2}\frac{\exp(x)}{(\exp(x)-1)^2} \, dx~,
\end{equation}
with the spin wave velocity $v_0\approx1.287\cdot10^5$~m/s \cite{Hayden91a-a} and $n_p=2/c$ the number of planes per unit length along the $c$-axis, which is the direction perpendicular to the planes ($c$ is the corresponding lattice constant). The integral is temperature dependent via its lower boundary $x_{\mathrm{0,i}}=\Delta_i/(k_BT)$, where $\Delta_{1}/k_B\approx26$~K and $\Delta_{2}/k_B\approx58$~K account for the anisotropy gaps which arise in $\rm La_2CuO4$ \cite{Keimer93}. However, the temperature dependence is weak in the $T$-range where the experimental data are discussed and one can approximate $\kappa_{\mathrm{mag}}\propto T^2$.

\subsection{Low-temperature characteristics -- thermal occupation}
% The investigation (cf. the fits in Fig.~\ref{low_t}) of the $T$-dependence of the data at low temperature reveals $\kappa_\mathrm{mag} \sim T$ for the 1D spin chain compound $\rm CaCu_2O_3$, $\kappa_\mathrm{mag} \sim T^2$ for the 2D-HAF and an exponential increase $\kappa_\mathrm{mag}\sim\exp (-\Delta/T)$, with $\Delta\approx 400$~K for the spin ladder compounds. This is exactly what is expected from the evaluation of the kinetic formula for $\kappa_\mathrm{mag}$ for these three cases \cite{Hess07,Hess03,Sologubenko00,Hess01} as long as the magnetic mean free path $l_\mathrm{mag}$ is temperature independent.
Interestingly, the experimental data shown in Fig.~\ref{low_t} exhibit extended regions at low $T$ where a reasonable description of the 
data with Eq.~\ref{kmag_simple} using a \textit{temperature independent} mean free path $l_\mathrm{mag}$ is possible. The solid 
lines in Fig.~\ref{low_t} represent fits where $l_\mathrm{mag}$ is a free, temperature independent 
parameter. A remarkably good description is found for $\kappa_\mathrm{mag}$ of the spin chain $\rm CaCu_2O_3$, as depicted in 
Fig.~\ref{low_t}a. Here, $\kappa_\mathrm{mag}$ is excellently described by a simple linear increase over the large temperature 
range $T\approx 100$-300~K and the fit yields $l_\mathrm{mag}=22 \pm 5$~{\AA} corresponding to about 5-6 lattice spacings.

For the spin ladder compounds $\rm (Sr,Ca,La)_{14}Cu_{24}O_{41}$ the situation is somewhat different.
%$\kappa_\mathrm{mag}$ can still nicely be fitted with Eq.~\ref{kappalad} 
As can be seen in Fig.~\ref{low_t}b, the temperature range over which a good fit with Eq.~\ref{kappalad} can be achieved is strongly reduced with respect to the previous case and apparently also depends on the composition of the material. In particular, the temperature interval where Eq.~\ref{kappalad} describes $\kappa_\mathrm{mag}$ with a $T$-independent mean free path is 54-102~K for $\rm Ca_9La_5Cu_{24}O_{41}$ but only 61-91~K for $\rm Sr_{14}Cu_{24}O_{41}$. Nevertheless, restricted to these $T$-ranges the fit (with the spin gap $\Delta$ and the mean free path $l_\mathrm{mag}$ as free fit parameters) yields similar results for both compounds. In particular, the value for the spin gap is found to be somewhat larger than but still in reasonable agreement with spin gap results from neutron scattering ($\Delta/k_B=418\pm15$~K and $\Delta/k_B=396\pm10$~K for $\rm Ca_9La_5Cu_{24}O_{41}$ and $\rm Sr_{14}Cu_{24}O_{41}$, respectively) \cite{Eccleston98,Katano99,Notbohm07}. The magnetic mean free path is of similar magnitude in both materials ($l_\mathrm{mag}=2980\pm110$~{\AA} and $l_\mathrm{mag}=2890\pm230\rm$~{\AA} respectively).\footnote{The slightly smaller values of $l_{\mathrm{mag}}$ and $\Delta$ for $\rm Ca_9La_5Cu_{24}O_{41}$ as compared to previous results \cite{Hess01} are a consequence of the usage of more accurate lattice parameters and an optimized fit-interval. It is stressed that these small corrections have no further consequences on the conclusions drawn in Ref.~\cite{Hess01}.} It is important to note that for temperatures higher than the mentioned ranges, Eq.~\ref{kappalad} completely fails to properly describe the experimental data with a $T$-independent $l_\mathrm{mag}$, as is also evident from Fig.~\ref{low_t}b. However, as will be discussed in Section~\ref{high_T}, a consistent picture arises if $l_\mathrm{mag}$ is allowed to become $T$-dependent at higher $T$. 
%The situation is different for temperatures below the discussed regime, since here $\kappa_\mathrm{mag}$ becomes negligible in comparison to $\kappa_\mathrm{ph}$ which prevents further analysis.

A similar observation is also made in the case of the 2D-HAF $\rm La_2CuO_4$ (cf. Fig.~\ref{low_t}c). Eq.~\ref{fit2d} with $l_\mathrm{mag}$ as a free parameter yields a reasonable fit in the range 70-158~K, where $l_\mathrm{mag}\approx558\pm140\rm$~{\AA} \cite{Hess03}. Again, the theoretical model fails to account for the high temperature regime as long as $l_\mathrm{mag}$ remains $T$-independent.

\subsection{Low temperature characteristics -- scattering off defects}

The actual meaning of the magnetic mean free path $l_\mathrm{mag}$ as a material parameter is \textit{a priori} not clear. On one hand it is known from the analogous case of phonon heat transport, that at low temperature the phonon mean free path can become as large as the crystal dimensions, i.e. of the order of millimeters \cite{Thacher67}. However, the extracted values for $l_\mathrm{mag}$ are several orders of magnitude smaller than the dimensions of the crystals which have been studied in the experiments. It therefore appears natural to conclude that $l_\mathrm{mag}$ should reflect the density of static defects in the material. On the other hand, a magnetic mean free path of the order of up to $\sim 1000$ lattice spacings implies almost perfect crystallinity of the underlying material, which is astonishing in view of the large and complicated unit cell of, for example, $\rm Ca_9La_5Cu_{24}O_{41}$. Note that a much smaller general magnitude of $l_\mathrm{mag}$ has been suggested for this compound based on Exact Diagonalization calculations of the thermal Drude weight which implied ballistic heat transport in spin ladder systems \cite{Alvarez02}. Such reduced values for $l_\mathrm{mag}$ could not be confirmed by more recent calculations which suggest a vanishing Drude weight \cite{Heidrich04,Zotos04}.

\begin{figure}
\resizebox{1\columnwidth}{!}{\includegraphics{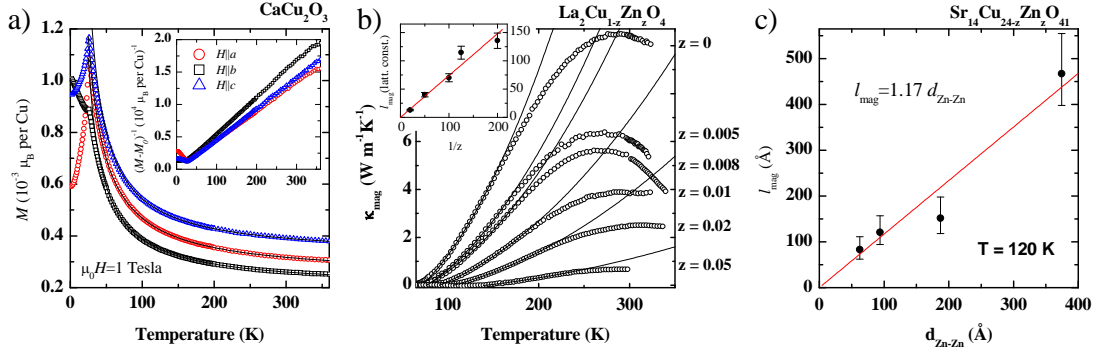} }
\caption{a) Magnetization of $\rm CaCu_2O_3$ as a function of temperature with a magnetic field $\mu_0H=1$~Tesla parallel to the three crystallographic axes. The solid lines represent Curie-Weiss-type fits to the data in the range 100-360~K which yield $\sim3$\% free spins with respect to Cu. Inset: Inverse magnetization after subtracting a constant $M_0$ which roughly accounts for van Vleck and chain magnetism. From Ref.~\cite{Hess07}. b) Main panel: Magnon thermal conductivity $\kappa_{\mathrm{mag}}$ of $\rm
La_2Cu_{1-z}Zn_zO_4$ ($z=0$, 0.005, 0.008, 0.01, 0.02, 0.05) as a function of $T$ (open circles).
Solid lines: fits according to Eq.~\ref{fit2d}. Inset: $l_{\mathrm{mag}}$ as a function of $1/z$ in
units of lattice constants $a$. Solid line: fit through origin. From Ref.~\cite{Hess03}. c) $l_{\mathrm{mag}}$ of $\rm Sr_{14}Cu_{24-z}Zn_zO_{41}$ ($z= 0.125$, 0.25, 0.5, 0.75)  as a function of the mean distance between Zn impurities $d_{\mathrm{Zn-Zn}}$. Solid line: linear fit through the origin. From Refs.~\cite{Hess07,Hess06,Hess03}.}
\label{lmagdat}
\end{figure}

A straightforward experimental method to elucidate the connection between $l_\mathrm{mag}$ and the density of magnetic defects in the material is to measure the latter independently from heat transport using a different experimental technique. One possibility is to study the magnetic susceptibility $\chi(T)$ where one has to assume that paramagnetic moments, the concentration of which is deducible from $\chi$, are connected with defects within the magnetic structure and hence may have an effect on the magnetic heat transport. For $\rm CaCu_2O_3$ the situation appears to be quite fortunate, since in a recent study Goiran et al. suggested a direct link between paramagnetic moments located off the 1D magnetic chain structures (detectable by $\chi$-measurements) and (possibly non-magnetic) defects within the chains \cite{Goiran06}. From susceptibility measurements on $\rm CaCu_2O_3$ (cf. Fig.~\ref{lmagdat}) it was then possible to deduce an upper limit for the mean distance between defects within a chain which turned out to be a factor of 2-3 larger than the extracted $l_\mathrm{mag}$ \cite{Hess07}. The same order of magnitude of both quantities indicates that these defects are the main scatterers for magnetic excitations within a chain. In the absence of a similar method for $\rm La_2CuO_4$ and $\rm (Sr,Ca,La)_{14}Cu_{24}O_{41}$, an alternative approach was selected to compare $l_\mathrm{mag}$ with known distances between intentionally doped defects in the material. Such defects can be induced by substituting a small amount of non-magnetic $\rm Zn^{2+}$ for the magnetic $\rm Cu^{2+}$-ions. This has been performed for both $\rm La_2CuO_4$ and $\rm Sr_{14}Cu_{24}O_{41}$ for a series of different doping levels. As can be inferred from Fig.\ref{lmagdat}b and \ref{lmagdat}c, once again $l_\mathrm{mag}$ has the same order of magnitude as the mean distance between the defects, which is in these cases well defined by the mean distance between the Zn dopants \cite{Hess03,Hess06}. This quantitatively confirms that a good analysis of $\kappa_\mathrm{mag}$ in these low dimensional spin systems can be performed using a rather simple kinetic model.

\subsection{Scattering processes at higher temperatures}
\label{high_T}
We now turn briefly to the behavior of $\kappa_\mathrm{mag}$ at high $T$ in order to elucidate the impact of temperature dependent scattering processes on the magnetic heat transport, i.e. scattering of magnetic excitations off other quasiparticles such as phonons, charge carriers or other magnetic excitations. Here we focus on heat transport in the spin ladder compounds, since these are susceptible to a variety of different types of doping. Most remarkable (apart from doping with non-magnetic impurities which has already been discussed) is certainly the possibility of hole-doping the ladders. Interestingly, the stoichiometric parent compound $\rm Sr_{14}Cu_{24}O_{41}$ is already inherently doped with holes. These holes reside only partially in the two-leg ladder structures -- the largest portion is located in chain substructures which are also present in the material beside the two-leg ladders. The holes are redistributed between the chain and ladder structures upon the isovalent substitution of Ca for Sr: with increasing Ca-content a significant increase of the hole concentration in the ladders is observed \cite{Osafune97,Nucker00,Rusydi07}. It is also possible to reduce the overall hole content in the material and thereby render the ladders virtually hole-free by replacing the divalent Sr or Ca by trivalent ions such as La \cite{Osafune97,Nucker00}.

A good example for the latter case of doping is the compound $\rm Ca_9La_5Cu_{24}O_{41}$ whose magnetic thermal conductivity is shown in Fig.~\ref{low_t}b in comparison with that of $\rm Sr_{14}Cu_{24}O_{41}$ \cite{Hess04a}. The effect of hole-doping on $\kappa_{\mathrm{mag}}$ can immediately be observed in this figure. For $T \lesssim100$~K the increase in $\kappa_{\mathrm{mag}}$ with $T$ is almost identical for both compounds. Pronounced differences only occur at higher $T$: $\kappa_{\mathrm{mag}}$ of $\rm La_5Ca_9Cu_{24}O_{41}$ exhibits a large peak ($\sim140~\rm Wm^{-1}K^{-1}$ at $\sim180$~K) and stays very large even at room temperature ($\sim100~\rm Wm^{-1}K^{-1}$). In contrast, the peak is much smaller in the case of $\rm Sr_{14}Cu_{24}O_{41}$ ($\sim75~\rm Wm^{-1}K^{-1}$ at $\sim150$~K). Here $\kappa_{\mathrm{mag}}$ decreases much more strongly at high $T$ and saturates at $\kappa_{\mathrm{mag}}\approx10~\rm Wm^{-1}K^{-1}$ for $T\gtrsim240$~K.

It is straightforward to attribute the strong high-$T$ suppression of $\kappa_{\mathrm{mag}}$ in $\rm Sr_{14}Cu_{24}O_{41}$  compared with $\rm La_5Ca_9Cu_{24}O_{41}$ to scattering of the magnons off holes, since the hole doping in this compound is the most relevant difference with respect to the undoped ladders of $\rm La_5Ca_9Cu_{24}O_{41}$. Both $\kappa_{\mathrm{mag}}$ curves are almost identical below a characteristic temperature $T_0\approx100$~K, confirming that this scattering mechanism becomes completely unimportant below $T_0$ and only reveals its full strength above a characteristic temperature $T^*\approx 240$~K. 

\begin{figure}
\resizebox{1\columnwidth}{!}{\includegraphics{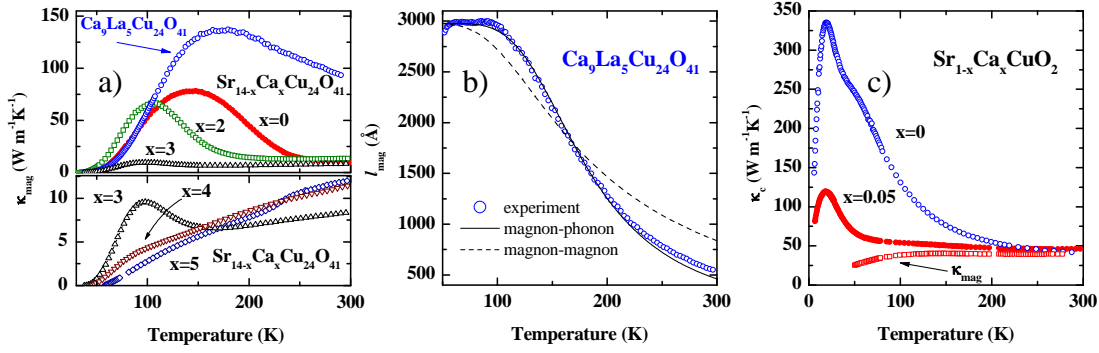}}
\caption{a) $\kappa_{\mathrm{mag}}(T)$ of $\rm Sr_{14-x}Ca_xCu_{24}O_{41}$ ($x=0$, 2, 3, 4,
5). Top panel: Data for ($x=0$, 2, 3) in comparison with $\kappa_{\mathrm{mag}}$ of $\rm La_5Ca_9Cu_{24}O_{41}$. Lower panel: enlarged representation for $x=3$, 4, 5. From Ref.~\cite{Hess04a}. b) $l_{\mathrm{mag}}$ of $\rm La_5Ca_9Cu_{24}O_{41}$ as a function of $T$. The solid and broken lines represent fits of $l_{\mathrm{mag}}$ accounting for magnon-phonon and magnon-magnon scattering respectively. From Ref.~\cite{Hess05}. c) The thermal conductivity along the chains $\kappa_c$ of Sr$_{1-x}$Ca$_x$CuO$_2$ at $x=0$ ($\circ$) and $x=0.05$ ($\bullet$). An estimation of $\kappa_{\mathrm{mag}}$ of the doped material via $\kappa_{\mathrm{mag}}\approx \kappa_\Vert - \kappa_\bot$ ($\kappa_\bot$ is not shown) is indicated. Data from Ref.~\cite{Ribeiro05}.}
\label{lmagscatter}
\end{figure}

The surprising temperature dependence of the strength of the magnon-hole scattering has been checked for robustness against changes of the hole content in the ladders, where $\kappa_{\mathrm{mag}}$ of a series of $\rm Sr_{14-x}Ca_xCu_{24}O_{41}$ single crystals has been investigated in detail \cite{Hess04a}. The comparison with $\kappa_{\mathrm{mag}}$ of the Ca-doped samples (see Fig.~\ref{lmagscatter}a) reveals that $T_0$ and $T^*$ are gradually shifted towards lower $T$; i.e., the temperature region where $\kappa_{\mathrm{mag}}$ is suppressed extends and magnon-hole-scattering also becomes important at low $T$. At $x=4$, 5 this region appears to becomes so wide that even the peak at low $T$ is suppressed. It was shown that the temperature dependence of $\kappa_{\mathrm{mag}}$ is unambiguously correlated with a charge ordered state in the compound, where charge ordering sets in at $T\lesssim T^*$. More precisely, the charge ordering in the ladders is accompanied by a drastic enhancement of the magnon mean free path $l_{\mathrm{mag}}$: the probability for magnon-hole scattering, which is close to unity for mobile holes, vanishes in the charge ordered state \cite{Hess04a}.

Turning to $\kappa_\mathrm{mag}$ of $\rm La_5Ca_9Cu_{24}O_{41}$, i.e. that of undoped ladders, it is \textit{a priori} clear that the magnon hole scattering, which is dominant in $\rm Sr_{14-x}Ca_xCu_{24}O_{41}$, cannot play any significant role. Nevertheless, the necessity to allow a formal $T$-dependence of $l_{\mathrm{mag}}$ at $T>100$~K indicates that further scattering processes are also relevant for this case. 
The only remaining possible scattering mechanism for magnons in this material are magnon-magnon or magnon-phonon scattering. In a more involved analysis the $T$-dependence of $l_{\mathrm{mag}}$ in $\rm La_5Ca_9Cu_{24}O_{41}$ has been calculated from the $\kappa_{\mathrm{mag}}$ data using Eq.~\ref{kappalad} by employing the previously extracted $\Delta=418$~K. The resulting $l_{\mathrm{mag}}(T)$ as shown in the main panel of Fig.~\ref{lmagscatter}b reflects the different $T$-regimes which govern $\kappa_\mathrm{mag}$. For $T\lesssim110$~K, $l_\mathrm{mag}$ is $T$-independent with a mean value $l_0=2980$~{\AA} which reflects the scattering of magnons off static defects. In order to describe the $T$-dependent $l_{\mathrm{mag}}$ at higher $T$ it was assumed that all scattering mechanisms were independent of each other and Matthiesen's rule applied: $l_{\mathrm{mag}}^{-1} = l_0^{-1}+\gamma_\mathrm{ph}d_\mathrm{ph}^{-1}+\gamma_\mathrm{mag}d_\mathrm{mag}^{-1}.$ 
Here $d_\mathrm{ph}$ and $d_\mathrm{mag}$ are the mean ''distances'' of phonons and magnons respectively, as calculated from the particle densities with $\gamma_\mathrm{ph}$ and $\gamma_\mathrm{mag}$ the corresponding scattering probabilities. Since it is unclear as to what extent the separate scattering mechanisms contribute to $l_{\mathrm{mag}}$, its behavior was analyzed based on the assumption that only one mechanism is active in addition to magnon-defect scattering.

The case of dominant magnon-phonon scattering was modelled by three energy-degenerate non-dispersive optical branches along the ladder direction, yielding $1/d_\mathrm{ph}=7.6\frac{1}{\exp(\Delta_\mathrm{opt}/T)-1}\,\cdot10^{9}\rm m^{-1}$ with $\Delta_\mathrm{opt}$ the optical gap (cf. Ref.\cite{Hess05} for details). The experimental $l_{\mathrm{mag}}$ was then fitted with $l_{\mathrm{mag}}^{-1} = l_0^{-1}+\gamma_\mathrm{ph}d_\mathrm{ph}^{-1}$
using $\gamma_\mathrm{ph}$ and $\Delta_\mathrm{opt}$ as free parameters. The fit (solid line in Fig.~\ref{lmagscatter}b) describes the data fairly well. Remarkably, the value found for $\Delta_\mathrm{opt}=795$~K is of the same order of magnitude as the energy of the longitudinal Cu-O stretching mode which is involved in the two-magnon-plus-phonon absorption observed in optical spectroscopy \cite{Gruninger00,Windt01}.

For the assumption of dominant magnon-magnon scattering, a less satisfactory agreement was obtained with $l_{\mathrm{mag}}^{-1} = l_0^{-1}+\gamma_\mathrm{mag}d_\mathrm{mag}^{-1}$, where $1/d_\mathrm{mag}=\frac{1}{\pi c_L}\int_0^\pi\frac{3}{3+\exp(\epsilon_k/k_BT)}dk$ (broken line in Fig.~\ref{lmagscatter}b). $c_L$ is the lattice constant along the ladders and $\epsilon_k$ was taken from Johnston et al. for the case of isotropic ladder coupling \cite{Johnston00}, with $\epsilon_{k=\pi}=\Delta=418$~K employed. 
The comparison between both fits suggests that scattering off optical phonons is dominant in this compound.

\section{Other developments}
\label{other}
Complementary to $\kappa_{\mathrm{mag}}$ of the ''dirty'' spin chains of $\rm CaCu_2O_3$, the spinon heat transport of the very ''clean'' spin chain materials $\rm Sr_2CuO_3$ and $\rm SrCuO_2$ has been investigated by A. V. Sologubenko and coworkers \cite{Sologubenko00a,Sologubenko01}.
Despite the data giving very clear evidence for spinon heat conductivity in this compounds, the precise extraction of $\kappa_{\mathrm{mag}}$ from the experimental data is quite involved, since the signature of $\kappa_{\mathrm{mag}}$ appears as a shoulder in the high-$T$ edge of the phononic low-$T$ peak (see Fig.~\ref{lmagscatter}c for the case of $\rm SrCuO_2$). In their analysis, Sologubenko et al. suggest that $\kappa_{\mathrm{mag}}$ exhibits a peak-like $T$-dependence which allows for further analysis \cite{Sologubenko00a,Sologubenko01}. In order to achieve a better extraction of $\kappa_{\mathrm{mag}}$, P. Ribeiro et al. investigated the thermal conductivity of $\rm Sr_{1-x}Ca_xCuO_2$ with $x=0.05$ with the expectation that the Ca-impurities selectively suppress $\kappa_{\mathrm{ph}}$ while $\kappa_{\mathrm{mag}}$ remains unchanged \cite{Ribeiro05}. However, as can be seen in Fig.~\ref{lmagscatter}c, the Ca-doping apparently leads to a complete suppression of the shoulder-like anomaly of the phonon peak and no signature of a peak structure remains (cf. Ref.~\cite{Ribeiro05} for details). Further studies are necessary to elucidate the origin of this intriguing observation.

\section{Conclusion}
From this summary of recent developments in the research on magnetic heat transport in quantum spin systems, it becomes evident that even though considerable progress has been made, this new thermal transport mechanism is still far from being thoroughly understood. 
We have seen that in some cases the magnetic heat conductivity can serve as a sensitive probe for magnetic excitations, which is a promising new approach to access the scattering mechanisms and dissipation of these excitations. However, many issues still remain to be resolved. For example, little is known about how $\kappa_{\mathrm{mag}}$ evolves when the magnetic systems become less quantum in nature, i.e. when $S>1/2$. Initial experiments and theoretical work have already addressed this topic \cite{Sologubenko03,Karadamoglou04,Savin05,Kordonis06} but the number of investigated materials of this type is still small. A better understanding of scattering processes (non-magnetic vs. magnetic impurities, phonons), frustration, and the effect of an external magnetic field will also be required. 

\begin{acknowledgement}
This overview was only made possible by the valuable contributions of P. Ribeiro, C. Baumann, H. ElHaes, A. Waske, U. Ammerahl, L. Colonescu, G. Krabbes, P. Reutler, A. Revcolevschi, G. Roth, C. Sekar, F. Heidrich-Meisner, W.Brenig, and  B. B\"{u}chner. It is a pleasure to further thank A. Chernyshev, J. Gemmer, A. Honecker, M. Michel, A. Rosch, A.V. Sologubenko, and X. Zotos for fruitful discussions, and A. P. Petrovic for proofreading the manuscript. This work was supported by the Deutsche Forschungsgemeinschaft through grant HE3439/7 and the European Commission through the FET open-STREP NOVMAG, Project Reference 032980.
\end{acknowledgement}
%
% % For tables use
% \begin{table}
% \caption{Please write your table caption here.}
% \label{tab:1}       % Give a unique label
% % For LaTeX tables use
% \begin{tabular}{lll}
% \hline\noalign{\smallskip}
% first & second & third  \\
% \noalign{\smallskip}\hline\noalign{\smallskip}
% number & number & number \\
% number & number & number \\
% \noalign{\smallskip}\hline
% \end{tabular}
% \end{table}
%
% \begin{center}
% % \begin{thebibliography}{}
% % % and use \bibitem to create references.
% % \bibitem{RefJ}
% % % Format for Journal Reference
% % Author, Journal \textbf{Volume}, (year) page numbers
% % % Format for books
% % \bibitem{RefB}
% % Author, \textit{Book title} (Publisher, place year) page numbers
% % % etc
% \end{center}
\bibliographystyle{epj}
\bibliography{kappamag}

\begin{thebibliography}{80}

\bibitem{Froehlich36}
H.~Fr{\"o}hlich, W.~Heitler, Proc. Roy. Soc. (London) \textbf{A155}, 640 (1936)

\bibitem{Luethi62}
B.~L{\"u}thi, J. Phys. Chem. Solids \textbf{23}, 35 (1962)

\bibitem{Douglass63}
R.L. Douglass, Phys. Rev. \textbf{129}, 1132 (1963)

\bibitem{Rives69}
J.E. Rives, G.S. Dixon, D.~Walton, J. Appl. Phys. \textbf{40}, 1555 (1969)

\bibitem{Walton73}
D.~Walton, J.E. Rives, Q.~Khalid, Phys. Rev. B \textbf{8}, 1210 (1973)

\bibitem{Berman}
R.~Berman, \emph{Thermal Conduction in Solids} (At the Clarendon Press, Oxford,
  1976)

\bibitem{Gorter69}
F.W. Gorter, L.J. Noordermeer, A.R. Kop, A.R. Miedema, Phys. Lett.
  \textbf{29A}, 331 (1969)

\bibitem{Lang77}
H.N.D. Lang, H.~van Kempen, P.~Wyder, Phys. Rev. Lett. \textbf{39}, 467 (1977)

\bibitem{Coenen77}
L.H.M. Coenen, H.N.D. Lang, J.H.M. Stoeling, H.~van Kempen, P.~Wyder, Physica B
  $\&$ C \textbf{86-88}, 968 (1977)

\bibitem{Hirakawa75}
H.~Miike, K.~Hirakawa, J. Phys. Soc. Japan \textbf{38}, 1279 (1975)

\bibitem{Zotos97}
X.~Zotos, F.~Naef, P.~Prelov{\v s}ek, Phys. Rev. B \textbf{55}, 11029 (1997)

\bibitem{Zotos99}
X.~Zotos, Phys. Rev. Lett. \textbf{82}, 1764 (1999)

\bibitem{Kudo99}
K.~Kudo, S.~Ishikawa, T.~Noji, T.~Adachi, Y.~Koike, K.~Maki, S.~Tsuji,
  K.~Kumagai, J. Low. Temp. Phys. \textbf{117}, 1689 (1999)

\bibitem{Sologubenko00}
A.V. Sologubenko, K.~Giann{\`o}, H.R. Ott, U.~Ammerahl, A.~Revcolevschi, Phys.
  Rev. Lett. \textbf{84}, 2714 (2000)

\bibitem{Hess01}
C.~Hess, C.~Baumann, U.~Ammerahl, B.~B{\"u}chner, F.~Heidrich-Meisner,
  W.~Brenig, A.~Revcolevschi, Phys. Rev. B \textbf{64}, 184305 (2001)

\bibitem{Hess02}
C.~Hess, U.~Ammerahl, C.~Baumann, B.~B{\"u}chner, A.~Revcolevschi, Physica B
  \textbf{312-313}, 612 (2002)

\bibitem{Hess03}
C.~Hess, B.~B{\"u}chner, U.~Ammerahl, L.~Colonescu, F.~Heidrich-Meisner,
  W.~Brenig, A.~Revcolevschi, Phys. Rev. Lett. \textbf{90}, 197002 (2003)

\bibitem{Hess04a}
C.~Hess, H.~{ElHaes}, B.~B{\"u}chner, U.~Ammerahl, M.~H{\"u}cker,
  A.~Revcolevschi, Phys. Rev. Lett. \textbf{93}, 027005 (2004)

\bibitem{Hess04}
{C. Hess and B. B\"uchner}, Eur. Phys. B \textbf{38}, 37 (2004)

\bibitem{Hess05}
C.~Hess, C.~Baumann, B.~B{\"u}chner, J. Mag. Mag. Mater. \textbf{290-291}, 322
  (2005)

\bibitem{Ribeiro05}
P.~Ribeiro, C.~Hess, P.~Reutler, G.~Roth, B.~B{\"u}chner, J. Mag. Mag. Mater.
  \textbf{290-291}, 334 (2005)

\bibitem{Hess06}
C.~Hess, P.~Ribeiro, B.~B{\"u}chner, H.~ElHaes, G.~Roth, U.~Ammerahl,
  A.~Revcolevschi, Phys. Rev. B \textbf{73}, 104407 (2006)

\bibitem{Hess07}
C.~Hess, H.~ElHaes, A.~Waske, B.~B{\"u}chner, C.~Sekar, G.~Krabbes,
  F.~Heidrich-Meisner, W.~Brenig, Phys. Rev. Lett. \textbf{98}, 027201 (2007)

\bibitem{Hess07a}
C.~Hess, B.~B{\"u}chner, J. Mag. Mag. Mater. \textbf{310}, e412 (2007)

\bibitem{Sologubenko00a}
A.V. Sologubenko, E.~Felder, K.~Giann{\`o}, H.R. Ott, A.~Vietkine,
  A.~Revcolevschi, Phys. Rev. B \textbf{62}, 6108 (2000)

\bibitem{Sologubenko01}
A.V. Sologubenko, K.~Giann{\`o}, H.R. Ott, A.~Vietkine, A.~Revcolevschi, Phys.
  Rev. B \textbf{64}, 054412 (2001)

\bibitem{Kudo01}
K.~Kudo, S.~Ishikawa, T.~Noji, T.~Adachi, Y.~Koike, K.~Maki, S.~Tsuji,
  K.~Kumagai, J. Phys. Soc. Jpn. \textbf{70}, 437 (2001)

\bibitem{Kudo03}
K.~Kudo, T.~Noji, Y.~Koike, T.~Nishizaki, N.~Kobayashi, J. Phys. Soc. Jpn.
  \textbf{72}, 2551 (2003)

\bibitem{Ando98}
Y.~Ando, J.~Takeya, D.L. Sisson, S.G. Doettinger, I.~Tanaka, R.S. Feigelson,
  A.~Kapitulnik, Phys. Rev. B \textbf{58}, 2913 (1998)

\bibitem{Vasilev98}
A.~Vasil'ev, V.~Pryadun, D.~Khomskii, G.~Dhalenne, A.~Revcolevschi, M.~Isobe,
  Y.~Ueda, Phys. Rev. Lett. \textbf{81}, 1949 (1998)

\bibitem{Hofmann02}
M.~Hofmann, T.~Lorenz, A.~Freimuth, G.S. Uhrig, H.~Kageyama, Y.~Ueda,
  G.~Dhalenne, A.~Revcolevschi, Physica B \textbf{312-313}, 597 (2002)

\bibitem{Hofmann03}
M.~Hofmann, T.~Lorenz, K.~Berggold, M.~Gr{\"u}ninger, A.~Freimuth, G.S. Uhrig,
  E.~Br{\"u}ck, Phys. Rev. B \textbf{67}, 184502 (2003)

\bibitem{Sologubenko03}
A.V. Sologubenko, S.M. Kazakov, H.R. Ott, T.~Asano, Y.~Ajiro, Phys. Rev. B
  \textbf{68}, 094432 (2003)

\bibitem{Sologubenko03a}
A.V. Sologubenko, H.R. Ott, G.~Dhalenne, A.~Revcolevschi, Europhys. Lett.
  \textbf{62}, 540 (2003)

\bibitem{Heidrich02}
F.~Heidrich-Meisner, A.~Honecker, D.C. Cabra, W.~Brenig, Phys. Rev. B
  \textbf{66}, 140406 (2002)

\bibitem{Heidrich03}
F.~Heidrich-Meisner, A.~Honecker, D.C. Cabra, W.~Brenig, Phys. Rev. B
  \textbf{68}, 134436 (2003)

\bibitem{Orignac03}
E.~Orignac, R.~Chitra, R.~Citro, Phys. Rev. B \textbf{67}, 134426 (2003)

\bibitem{Alvarez02}
J.V. Alvarez, C.~Gros, Phys. Rev. Lett. \textbf{89}, 156603 (2002)

\bibitem{Heidrich04}
F.~Heidrich-Meisner, A.~Honecker, D.C. Cabra, W.~Brenig, Phys. Rev. Lett.
  \textbf{92}, 069703 (2004)

\bibitem{Gros04}
C.~Gros, J.V. Alvarez, Phys. Rev. Lett. \textbf{92}, 069704 (2004)

\bibitem{Saito03}
K.~Saito, Phys. Rev. B \textbf{67}, 064410 (2003)

\bibitem{Saito03a}
K.~Saito, Europhys. Lett. \textbf{61}, 34 (2003)

\bibitem{Shimshoni03}
E.~Shimshoni, N.~Andrei, A.~Rosch, Phys. Rev. B \textbf{68}, 104401 (2003)

\bibitem{Klumper02}
A.~Kl{\"u}mper, K.~Sakai, J. Phys. A: Math. Gen. \textbf{35}, 2173 (2002)

\bibitem{Sakai03}
K.~Sakai, A.~Kl{\"u}mper, J.Phys. A \textbf{36}, 11617 (2003)

\bibitem{Zotos04}
X.~Zotos, Phys. Rev. Lett. \textbf{92}, 067202 (2004)

\bibitem{Karadamoglou04}
J.~Karadamoglou, X.~Zotos, Phys. Rev. Lett. \textbf{93}, 177203 (2004)

\bibitem{Prelovsek04}
P.~Prelovsek, S.E. Shawish, X.~Zotos, M.~Long, Phys. Rev. B \textbf{70}, 205129
  (2004)

\bibitem{Louis06}
K.~Louis, P.~Prelovsek, X.~Zotos, Phys. Rev. B \textbf{74}, 235118 (2006)

\bibitem{Li02}
M.R. Li, E.~Orignac, Europhys. Lett. \textbf{60}, 432 (2002)

\bibitem{Rozhkov05}
A.V. Rozhkov, A.L. Chernyshev, Phys. Rev. Lett. \textbf{94}, 087201 (2005)

\bibitem{Chernyshev05}
{A. L. Chernyshev and A. V. Rozhkov}, Phys. Rev. B \textbf{72}, 104423 (2005)

\bibitem{Jung06}
{P. Jung and R.W. Helmes and A. Rosch}, Phys. Rev. Lett. \textbf{96}, 067202
  (2006)

\bibitem{Kordonis06}
K.~Kordonis, A.V. Sologubenko, T.~Lorenz, S.W. Cheong, A.~Freimuth, Phys. Rev.
  Lett. \textbf{97}, 115901 (2006)

\bibitem{Boulat06}
E.~Boulat, P.~Metha, N.~Andrei, E.~Shimshoni, A.~Rosch,
  arXiv:cond-mat/0607837v1  (unpublished)

\bibitem{Kiryukhin01}
V.~Kiryukhin, Y.J. Kim, K.J. Thomas, F.C. Chou, R.W. Erwin, Q.~Huang, M.A.
  Kastner, R.J. Birgeneau, Phys. Rev. B \textbf{63}, 144418 (2001)

\bibitem{Motoyama96}
N.~Motoyama, H.~Eisaki, S.~Uchida, Phys. Rev. Lett. \textbf{76}(17), 3212
  (1996)

\bibitem{Dagotto99}
E.~Dagotto, Rep. Prog. Phys. \textbf{62}, 1525 (1999)

\bibitem{Kluemper93}
A.~Kl{\"u}mper, Z. Phys. B \textbf{91}, 507 (1993)

\bibitem{Faddeev81}
L.~Faddeev, L.~Takhtajan, Phys. Lett. A \textbf{85}, 375 (1981)

\bibitem{Manousakis91}
E.~Manousakis, Rev. Mod. Phys. \textbf{63}, 1 (1991)

\bibitem{Coldea01}
R.~Coldea, S.M. Hayden, G.~Aeppli, T.G. Perring, C.D. Frost, T.E. Mason, S.W.
  Cheong, Z.~Fisk, Phys. Rev. Lett. \textbf{86}, 5377 (2001)

\bibitem{Sandvik01}
A.W. Sandvik, R.R.P. Singh, Phys. Rev. Lett. \textbf{86}, 528 (2001)

\bibitem{Ho01}
C.M. Ho, V.N. Muthukumar, M.~Ogata, P.W. Anderson, Phys. Rev. Lett.
  \textbf{86}, 1626 (2001)

\bibitem{Dagotto92}
E.~Dagotto, J.~Riera, D.~Scalapino, Phys. Rev. B \textbf{45}, 5744 (1992)

\bibitem{Dagotto96}
E.~Dagotto, T.M. Rice, Science \textbf{271}, 618 (1996)

\bibitem{Hayden91a-a}
S.M. Hayden, G.~Aeppli, R.~Osborn, A.D. Taylor, T.G. Perring, S.W. Cheong,
  Z.~Fisk, Phys. Rev. Lett. \textbf{67}, 3622 (1991)

\bibitem{Keimer93}
B.~Keimer, R.J. Birgeneau, A.~Cassanho, Y.~Endoh, M.~Greven, M.A. Kastner,
  G.~Shirane, Z. Phys. B \textbf{91}, 373 (1993)

\bibitem{Eccleston98}
R.S. Eccleston, M.~Uehara, J.~Akimitsu, H.~Eisaki, N.~Motoyama, S.~Uchida,
  Phys. Rev. Lett. \textbf{81}, 1702 (1998)

\bibitem{Katano99}
S.~Katano, T.~Nagata, J.~Akimitsu, M.~Nishi, K.~Kakurai, Phys. Rev. Lett.
  \textbf{82}, 636 (1999)

\bibitem{Notbohm07}
S.~Notbohm, P.~Ribeiro, B.~Lake, D.A. Tennant, K.P. Schmidt, G.S. Uhrig,
  C.~Hess, R.~Klingeler, G.~Behr, B.~B{\"u}chner et~al., Phys. Rev. Lett.
  \textbf{98}, 027403 (2007)

\bibitem{Thacher67}
P.D. Thacher, Physical Review \textbf{156}, 975 (1967)

\bibitem{Goiran06}
M.~Goiran, M.~Costes, J.M. Broto, F.C. Chou, E.~Arushanov, S.L. Drechsler,
  B.~B{\"u}chner, V.~Kataev, New J. Phys. \textbf{8}, 74 (2006)

\bibitem{Osafune97}
T.~Osafune, N.~Motoyama, H.~Eisaki, S.~Uchida, Phys. Rev. Lett. \textbf{78},
  1980 (1997)

\bibitem{Nucker00}
N.~N{\"u}cker, M.~Merz, C.A. Kuntscher, S.~Gerhold, S.~Schuppler, R.~Neudert,
  M.S. Golden, J.~Fink, D.~Schild, S.~Stadler et~al., Phys. Rev. B \textbf{62},
  14384 (2000)

\bibitem{Rusydi07}
A.~Rusydi, M.~Berciu, P.~Abbamonte, S.~Smadici, H.~Eisaki, Y.~Fujimaki,
  S.~Uchida, M.~Rubhausen, G.A. Sawatzky, Phys. Rev. B \textbf{75}, 104510
  (2007)

\bibitem{Gruninger00}
M.~Gr{\"u}ninger, D.~{van der Marel}, A.~Damascelli, A.~Erb, T.~Nunnner,
  T.~Kopp, Phys. Rev. B \textbf{62}, 12422 (2000)

\bibitem{Windt01}
M.~Windt, M.~Gr{\"u}ninger, T.~Nunner, C.~Knetter, K.P. Schmidt, G.S. Uhrig,
  T.~Kopp, A.~Freimuth, U.~Ammerahl, B.~B{\"u}chner et~al., Phys. Rev. Lett.
  \textbf{87}, 127002 (2001)

\bibitem{Johnston00}
D.C. Johnston, M.~Troyer, S.~Miyahara, D.~Lidsky, K.~Ueda, M.~Azuma, Z.~Hiroi,
  M.~Takano, M.~Isobe, Y.~Ueda et~al., arXiv:cond-mat/0001147v1  (unpublished)

\bibitem{Savin05}
A.V. Savin, G.P. Tsironis, X.~Zotos, Phys. Rev. B \textbf{72}, 140402 (2005)

\end{thebibliography}
% 
% \end{thebibliography}
% 
\end{document}